\documentclass[pre,aps,twocolumn,superscriptaddress,showpacs]{revtex4}%{revtex4-1}
\usepackage{amssymb,epsfig,amsmath,bm,subfigure,graphicx,textcomp,url,float,color,cases}

\begin{document}
\title{Macroscopic fluctuation theory and first-passage properties of surface diffusion}

\author{Baruch Meerson}
\email{meerson@mail.huji.ac.il}
\affiliation{Racah Institute of Physics, Hebrew
  University of Jerusalem, Jerusalem 91904, Israel}

\author{Arkady Vilenkin}
\email{vilenkin@mail.huji.ac.il}
\affiliation{Racah Institute of Physics, Hebrew
  University of Jerusalem, Jerusalem 91904, Israel}


\begin{abstract}

We investigate non-equilibrium fluctuations of a solid surface governed by the stochastic Mullins-Herring equation with conserved noise. This equation describes surface diffusion of adatoms accompanied by their exchange between the surface and the bulk of the solid,
when desorption of adatoms is negligible. Previous works dealt with dynamic scaling behavior of the fluctuating interface. Here we determine the probability that the interface first reaches a large given height at a specified time. We also find the %most likely
optimal time history of the interface conditional on this non-equilibrium fluctuation. We obtain these results by developing a  macroscopic fluctuation theory of surface diffusion.

\end{abstract}

\maketitle
The stochastic Mullins-Herring equation with conserved noise can be written in a rescaled form as
\begin{equation}\label{mlnns}
\partial_{t}h=-\nabla^4 h+\nabla \cdot \bm{\xi}(\mathbf{x},t),
\end{equation}
where $h(\mathbf{x},t)$ is the interface height, and $\bm{\xi}$ is a Gaussian
noise which is delta-correlated in $\mathbf{x}$ and $t$ and has zero mean and unit variance.

The noiseless version of Eq.~(\ref{mlnns}) was suggested by Mullins almost 60 years ago \cite{Mullins}.
It describes the capillary flattening of a perturbed solid surface to its equilibrium
shape, where the surface diffusion of adatoms is accompanied by adatom
exchange between the surface and the bulk of the solid, while the adatom desorption is negligible
\cite{Villain1,Villain2,Zangwill,Cahn,Vilenkin}. Equation~(\ref{mlnns}) was extensively studied in the context of dynamic scaling behavior of fluctuating interfaces \cite{Siegert,Racz,Barabasi,Krug}.  A closely related discrete model
is the restricted solid-on-solid
model with conserved noise \cite{Kim}. In this model a pair of nearest-neighbor sites $(i, i+1)$ is randomly selected on a $(d-1)$-dimensional substrate.
One particle is moved from the site $i$ to the site $i+1$, or vice versa, with probability $1/2$. The move, however, is only allowed if
the inequality $|h(i+1,t)-h(i,t)|\leq 2$ is satisfied after the move.

The previous works \cite{Siegert,Racz,Kim} focused on
the interface width $w(L,t)$ which, for the conserved equation like Eq.~(\ref{mlnns}), is defined by
$w^2(L,t)=\langle h^2(\mathbf{r},t)\rangle$, where $L$ is the linear size of the system.
Assuming an initially flat interface,  $w(L,t)$ exhibits scaling behavior:
\begin{numcases}
{w(L,t) \sim}
t^{\beta}\,,   & \mbox{$0\ll t \ll L^4$}, \label{growthexp}\\
L^{\alpha}\,, \!\!\!\!\!\!&\mbox{$t\gg L^4$}, \label{eqexp}
\end{numcases}
where $\beta=(2-d)/8$, $\alpha=(2-d)/2$, and $d$ is the dimension of space  \cite{Racz,Siegert,Barabasi,Krug,Kim}. The dynamic exponent $z=4$ is related to $\alpha$ and $\beta$ via $z=\alpha/\beta$.  The most interesting case here is $d=1$, when both the growth exponent $\beta=1/8$, and the roughness exponent $\alpha=1/2$ are positive. That is, in one dimension the small  intrinsic noise makes the interface roughen without any external driving, in spite of the smoothing effect of the surface diffusion. The roughening disappears if one replaces in Eq.~(\ref{mlnns}) the surface diffusion term $-\nabla^4 h$ by the more customary term  $\nabla^2 h$ \cite{Siegert,Family,Barabasi,Krug}.

The interface width $w(L,t)$ is a useful \emph{integral} measure of the fluctuating interface.
Here we will address Eq.~(\ref{mlnns}) from a different angle, by focusing on a \emph{first-passage} property
of the interface \cite{Brayetal}. We will evaluate the probability that the interface first reaches a (large) given height at time $T$, and find the optimal history of the interface conditional on this non-equilibrium event.

To achieve this goal we will develop a macroscopic fluctuation theory (MFT) for
Eq.~(\ref{mlnns}). This is a WKB-like theory in the spirit of the (weak-noise limit of) Martin-Siggia-Rose field-theoretical formalism \cite{MSR}. More recently, variants of this theory were developed for diffusive lattice gases: for their non-equilibrium steady states, see \cite{MFTreview} for a review, and for non-stationary settings \cite{nonstationary1,nonstationary2,nonstationary3}.  Using the MFT formulation, we will first evaluate the probability to observe, at $t=T$,
a specified height profile $h_T(x)$ when starting from a flat profile at $t=0$. The solution also gives the optimal
height profile history leading to this $h_T(x)$. Then we will deal with the first-passage problem. Here there are two different regimes: the equilibrium regime at $T\gg L^4$, and the non-equilibrium regime at $T\ll L^4$. In the equilibrium regime the first-passage probability can be found from a minimization of the free energy of the system under proper constraints, whereas the optimal activation history coincides with the time-reversed relaxation history, as to be expected from the Onsager-Machlup symmetry \cite{Onsager}. In the non-equilibrium regime a full time-dependent solution of the MFT equations is needed for the evaluation of the first-passage probability, whereas the activation and time-reversed relaxation histories are entirely different.
In the non-equilibrium regime we uncover dynamic scaling behavior of the first-passage probability and of the height profile corresponding to reaching the maximum height at $t=T$. Finally, we will show how to extend some of our results to the Mullins-Herring equation with non-conserved noise.

Let us measure the coordinate $x$ along the substrate, $|x|\leq L/2$, and assume periodic boundaries. The MFT equations follow from a  saddle-point evaluation of the action integral, obtained in a standard way from Eq.~(\ref{mlnns}) \cite{SM}. They can be written as two Hamiltonian partial differential equations:
\begin{eqnarray}
  \partial_{t}h &=&\delta H/\delta p=-\partial_{x}(\partial_{x}^3h+\partial_{x}p), \label{eqh} \\
   \partial_{t}p &=&-\delta H/\delta h= \partial_{x}^4p,\label{eqp}
\end{eqnarray}
%These equations are Hamiltonian, $\partial_t h = \delta H/\delta p$ and $\partial_t p = -\delta H/\delta h$,
with the Hamiltonian
\begin{equation}\label{H1}
H=\int_{-L/2}^{L/2}dx\,\mathcal{H},\quad \mathcal{H}
= \partial_x p \left(\partial_{x}^{3}h+\frac{1}{2}\partial_{x}p\right).
\end{equation}
With proper initial and boundary conditions, that we will discuss shortly, Eqs.~(\ref{eqh}) and (\ref{eqp}) describe the optimal interface height history $h(x,t)$. Once they are solved, we can evaluate the probability
$\mathcal{P}$ of the large deviation we are interested in: $-\ln {\mathcal P} \simeq S$, where
\begin{eqnarray}
  S &=& \int_{0}^{T}dt\int_{-L/2}^{L/2}dx\,(p\partial_t h - \mathcal{H}) \nonumber \\
  &=& \frac{1}{2}\int_{0}^{T}dt\int_{-L/2}^{L/2}dx\,(\partial_{x}p)^{2} \label{action}
\end{eqnarray}
is the action. Let us briefly discuss some general properties of this Hamiltonian flow.
As follows from Eq.~(\ref{H1}), there are two invariant zero-energy manifolds. The manifold $\partial_x p=0$ corresponds to the noiseless Mullins-Herring equation $\partial_t h = -\partial_x^4 h$.
The second invariant zero-energy manifold,  described by the equation
\begin{equation}\label{eqmanifold}
\partial_x p=-2\partial_x^3 h ,
\end{equation}
corresponds to thermal equilibrium. Using Eqs.~(\ref{eqh}) and (\ref{eqmanifold}), we obtain
\begin{equation}\label{timereversed}
\partial_t h = \partial_x^4 h.
\end{equation}
That is, the equilibrium dynamics is described by the \emph{time-reversed} Mullins-Herring equation. As a result, an activation trajectory at thermal equilibrium coincides with the time-reversed relaxation trajectory \cite{Onsager}.

Before dealing with the first-passage problem, we will solve an auxiliary problem by specifying a height profile $h_T(x)$ at time $t=T$. For simplicity, we assume a flat interface at $t=0$ and set $h(x,t=0)=0$.  In the limit of $T\gg L^4$, the system will explore equilibrium fluctuations in order to reach $h_T(x)$. In this limit
there is no need to find the activation trajectory: It suffices to evaluate the difference between the free energies of the final and initial states. Indeed, let us evaluate the action (\ref{action}) on the equilibrium manifold~(\ref{eqmanifold}), using Eq.~(\ref{timereversed}):
\begin{eqnarray}
  \!\!\!\!\!\!S &=&  2 \int dt\int dx \,(\partial_x^3 h)^2  = 2 \int dt \int dx \,\partial_x^5 h\,\partial_x h \nonumber\\
   \!\!\!\!\!\!&=&  2 \int dt \int dx \,\partial_{xt}^2 h\,\partial_x h =
    \int_{-L/2}^{L/2}dx \,[\partial_x h_T(x)]^2 \label{freen}.
\end{eqnarray}
The final expression is the free energy cost of the height profile $h_T(x)$, as to be expected.

For finite $T$ the system is out of equilibrium, and we need to solve Eqs.~(\ref{eqh}) and (\ref{eqp}) explicitly.
We can write
\begin{eqnarray}
  h(x,t) &=& \sum_{n=1}^{\infty} a_n(t) \cos (k_n x) + b_n(t) \sin (k_n x), \label{hform}\\
  p(x,t) &=& \sum_{n=1}^{\infty} \alpha_n(t) \cos (k_n x) + \beta_n(t) \sin (k_n x), \label{pform}
\end{eqnarray}
where $k_n=2\pi n/L$. Solving the ensuing ordinary differential equations for the coefficients,
we obtain
\begin{eqnarray}
  \left[ \begin{array}{c} a_n(t) \\ b_n(t) \end{array} \right] &=&
  \left[ \begin{array}{c} A_n \\ B_n \end{array} \right] \times \frac{\sinh (k_n^4 t)}{\sinh (k_n^4 T)}, \quad\mbox{and}\label{anbn} \\
  \left[ \begin{array}{c} \alpha_n(t) \\ \beta_n(t) \end{array} \right] &=& \left[ \begin{array}{c} A_n \\ B_n \end{array} \right] \times \frac{k_n^2 e^{k_n^4 t}}{\sinh (k_n^4 T)}, \label{alphanbetan}
\end{eqnarray}
%\begin{equation}\label{anbnold}
%\left[ \begin{array}{c} a_n(t) \\ b_n(t) \end{array} \right] = \left[ \begin{array}{c} A_n \\ B_n \end{array} \right] \times \frac{\sinh (k_n^4 t)}{\sinh (k_n^4 T)},
%\end{equation}
%and
%\begin{equation}\label{alphanbetanold}
%\left[ \begin{array}{c} \alpha_n(t) \\ \beta_n(t) \end{array} \right] = \left[ \begin{array}{c} A_n \\ B_n \end{array} \right] \times \frac{k_n^2 e^{k_n^4 t}}{\sinh (k_n^4 T)},
%\end{equation}
where $A_n$ and $B_n$ are the Fourier coefficients of $h_T(x)$. Now we evaluate the double integral in Eq.~(\ref{action}) and obtain
\begin{equation}
   S = L\sum_{n=1}^{\infty}K_{n}(L,T)\left(A_{n}^{2}+B_{n}^{2}\right), \label{acn2}
\end{equation}
where
\begin{equation}\label{Kn}
 K_n(L,T)=\frac{k_n^2}{2(1-e^{-2k_n^{4}T})}.
\end{equation}
Equations~(\ref{acn2}) and (\ref{Kn}) yield the probability of observing a given height profile $h_T(x)$ at an arbitrary time $T$. This is a simple but fully non-equilibrium result.  %This simple but fully non-equilibrium result owes its existence to the fact that Eqs.~(\ref{eqh}) and (\ref{eqp}) are linear and easily solvable.

For $T \gg L^4$  $K_n(L,T)\simeq k_n^2/2$, and Eq.~(\ref{acn2}) reduces to the free energy from Eq.~(\ref{freen}), whereas
the activation trajectory coincides with the time-reversed relaxation trajectory.  For $T\ll L^4$ the system
is far from equilibrium.  As a simple example, suppose that $h_T(x)$ only includes
long wavelengths, comparable to $L$. Then
$$
S\simeq\frac{L}{4T}\sum_{n=1}^{n_{max}}\frac{A_{n}^{2}+B_{n}^{2}}{k_n^2},
$$
which diverges at $T\to 0$ implying a very small probability of observing a large given height at short time. In their turn, the Fourier coefficients $a_n$ and $b_n$ from  Eq.~(\ref{anbn}) can be approximated as $a_{n}(t)\simeq(t/T) A_n$ and $b_{n}(t)\simeq(t/T) A_n$, so
that the optimal interface height history becomes simply $h(x,t)\simeq (t/T) h(x,T)$. This is very different from the time-reversed relaxation history at equilibrium.

\begin{figure}
\includegraphics[width=0.4\textwidth,clip=]{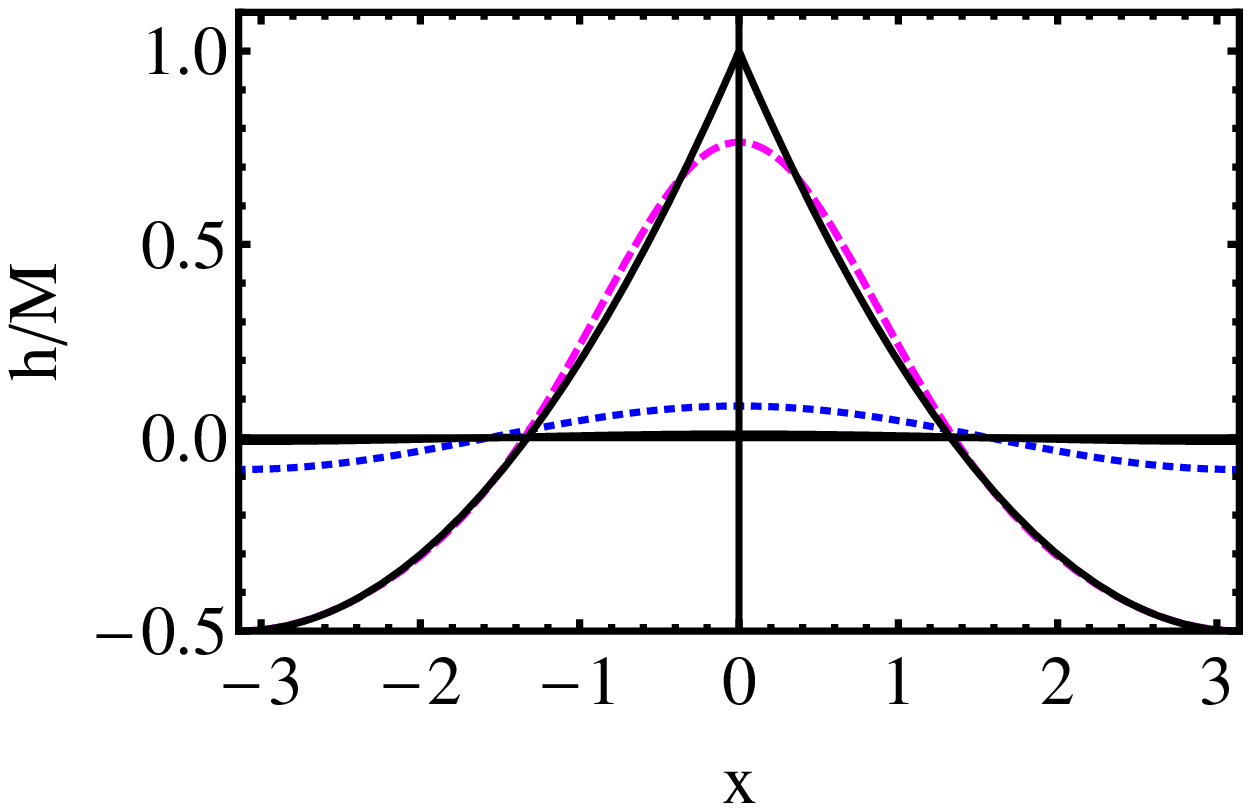}
\includegraphics[width=0.4\textwidth,clip=]{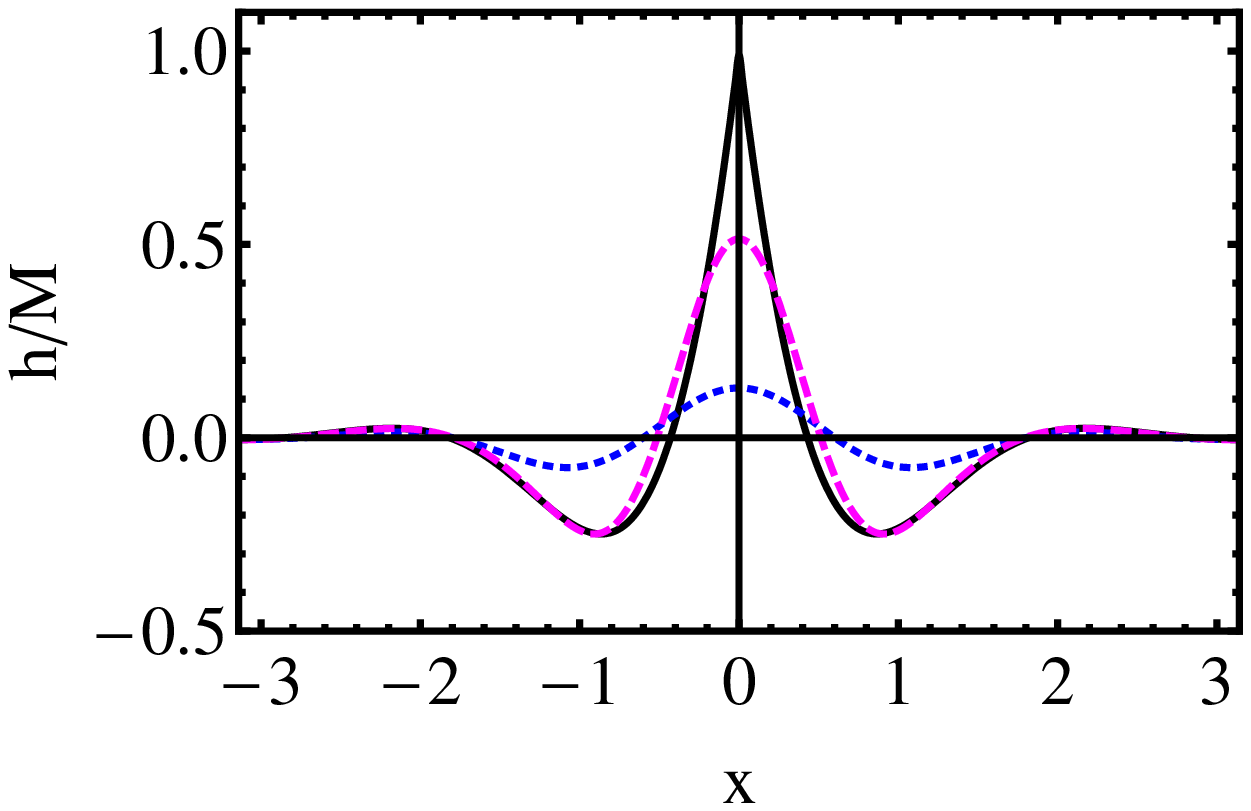}
\caption{(Color online) The optimal interface height history conditional on reaching $h=M$ at $t=T$. The initial state is $h(x,t=0)=0$. Shown is the rescaled height $h(x,t)/M$ versus the rescaled coordinate $2\pi x/L$.  Top panel: $T=5$ and $t=3$ (dotted line), $4.99$ (dashed line), and $t=5$ (solid line).  %, the action here is
%$s=LM^2/3.28978$.
Bottom panel: $T=0.01$ and t=$0.003$ (dotted line),  $0.009$ (dashed line), and $t=0.01$ (solid line). %Value of the action [see Eq.(\ref{acn2}) is $s=LM^2/0.921671$.
Time is rescaled by $(L/2\pi)^4$.}
\label{history}
\end{figure}

With Eqs.~(\ref{hform})-(\ref{Kn}) at hand, we can now evaluate the probability that the interface first reaches a (large) given height, say $M$, at time $T$. Without losing generality, we assume that
this happens at $x=0$. We start from  minimization of the action (\ref{acn2}) under constraint $h_T(x=0)=M$, without demanding that this height was not reached at earlier times $0<t<T$. In view of Eq.~(\ref{hform}), the constraint $h_T(x=0)=M$
becomes
\begin{equation}\label{hx*}
\sum_{n=1}^{\infty}A_{n} = M.
\end{equation}
Let us introduce constrained action,
$$
S_{\lambda}= \sum_{n=1}^{\infty} \left[L K_{n}(L,T) (A_{n}^{2}+B_{n}^{2}) -\lambda A_n\right],
$$
where $\lambda$ is a Lagrange multiplier. Varying $S_{\lambda}$ with respect to $A_n$ and $B_n$, we obtain
$$
\sum_{n=1}^{\infty} \left[ 2 L K_{n}(L,T) (A_{n}\delta A_{n}+B_{n}\delta B_{n})-
\lambda \delta A_{n}\right]=0,
$$
which yields
\begin{equation*}
A_{n}=\frac{\lambda }{2LK_{n}(L,T)},\quad B_{n}=0,\quad n=1,2,\dots .
\end{equation*}
Plugging these $A_n$ into Eq.~(\ref{hx*}), we find
\begin{equation}\label{lambda}
\lambda=\frac{2LM}{\sum_{n=1}^{\infty}K_{n}^{-1}(L,T)}=\frac{LM}{Q(L,T)},
\end{equation}
where
\begin{equation}\label{QT}
Q(L,T)=\sum_{n=1}^{\infty}\frac{1-e^{-2k_n^{4} T}}{k_n^{2}}\,.
\end{equation}
Now we use Eq.~(\ref{anbn}) and find the time-dependent Fourier coefficients $a_n$ and $b_n$:
\begin{equation}\label{anoft}
a_{n}(t)=\frac{M \sinh(k_n^{4} t)}{2K_{n}(T)Q(L,T)\sinh(k_n^{4} T)},\quad b_{n}(t)=0.
\end{equation}
Therefore,  the optimal height profile history is
\begin{equation}\label{optimal}
h(x,t)=\frac{M}{Q(L,T)}\sum_{n=1}^{\infty} \frac{1-e^{-2k_n^4T}}{k_n^2}\,\frac{\sinh(k_n^4 t)}{\sinh(k_n^4 T)} \, \cos (k_n x).
\end{equation}
Importantly, the value $h=M$ is achieved at $t=T$ for the first time. Therefore, Eq.~(\ref{optimal}) solves the first-passage problem that we are after. Figure \ref{history} shows the optimal interface height histories $h(x,t)$ for large (the top panel) and small (the bottom panel) values of $T$. As one can see, these histories are very different. This includes the height profiles at $t=T$,
\begin{equation}\label{atT}
h(x,T)=\frac{M}{Q(L,T)}\sum_{n=1}^{\infty} \frac{1-e^{-2k_n^4 T}}{k_n^2}\, \cos (k_n x).
\end{equation}
which, surprisingly, develop a corner singularity at $x=0$ where the value $h=M$ is first reached. Note also that at large $T$ (the top panel) most of the dynamics happens at times close to $T$: the system has enough time to
thermalize before producing a large deviation.

The first-passage action can be found from Eq.~(\ref{acn2}):
\begin{equation}\label{acn3}
-\ln \mathcal{P}\simeq S=\frac{LM^{2}}{2Q(L,T)}.
\end{equation}
The resulting probability $\mathcal{P}(M)$ is Gaussian, with the standard deviation  $\sigma(L,T)=\sqrt{Q(L,T)/L}$. %Let us consider two asymptotic regimes: the equilibrium regime at $T\gg L^4$, and the non-equilibrium regime at $T\ll L^4$.
The asymptotics of the function $Q(L,T)$ are the following:
\begin{numcases}
{Q(L,T) \simeq}
L^2/24 \,,   & \mbox{$T\gg L^4$}, \label{Qlong}\\
\frac{\Gamma(3/4) \,T^{1/4}\,L}{2^{3/4}\pi}\,, &\mbox{$T\ll L^4$},\label{Qshort}
\end{numcases}
where $\Gamma(\dots)$ is the gamma-function. The $T\gg L^4$  asymptotic is obtained by neglecting $e^{-2k_n^4 T}$ compared to $1$; the remaining infinite sum is equal to $L^2/24$.
To obtain the $T\ll L^4$ asymptotic, we replaced the infinite sum in Eq.~(\ref{QT}) by the integral
which yields Eq.~(\ref{Qshort}). The standard deviation of  $\mathcal{P}(M)$ has the following asymptotics:
\begin{numcases}
{\sigma(L,T) \simeq}
\frac{L^{1/2}}{2\sqrt{6}}\,,   & \mbox{$T\gg L^4$}, \label{Vlong}\\
\frac{[\Gamma(3/4)]^{1/2} \,T^{1/8}}{2^{3/8} \pi^{1/2}}\,,&\mbox{$T\ll L^4$}. \label{Vshort}
\end{numcases}
At long times $\sigma$ is independent of $T$, and its scaling with $L$ is described by the roughness exponent $\alpha=1/2$.
At short times $\sigma$ does not depend on $L$, and its scaling with $T$ exhibits the growth exponent $\beta=1/8$.

\begin{figure}
\includegraphics[width=0.4\textwidth,clip=]{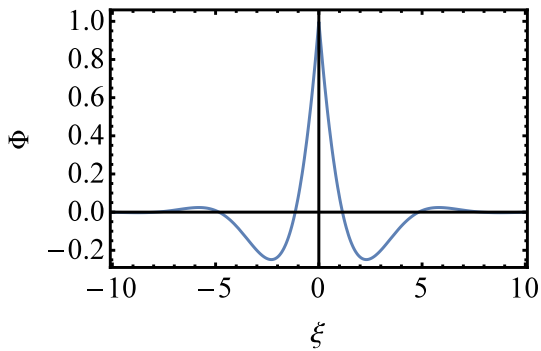}
\caption{(Color online.) The function $\Phi(\xi)=h(x,T)/M$ from Eq.~(\ref{Phiresult}).}
\label{Phi}
\end{figure}

Now let us examine the optimal height profiles at $t=T$ in the long- and short-time limits. In the long-time limit
\begin{eqnarray}
   h(x,T\to \infty) &=& \frac{6 M}{\pi^2}\sum_{n=1}^{\infty}\frac{\cos (k_nx)}{n^2} \nonumber \\
 &=& M\left[6\left(\frac{x}{L}\right)^2 - \frac{6\left|x\right|}{L}+1 \right]. \label{hTlong}
\end{eqnarray}
This shape is very close to that observed at $t=T$ on the top panel of Fig.~\ref{history}. In the short-time limit we can replace the infinite sum in Eq.~(\ref{atT}) by an integral. After a change of variable,
\begin{equation}\label{atT1}
h(x,T\ll 1)\simeq \frac{M}{\Gamma(3/4)}\,\int_0^{\infty} \frac{dz}{z^2} \left(1-e^{-z^4}\right)\, \cos (\xi z),
\end{equation}
where $\xi=x/(2T)^{1/4}$. The profile $h(x,T)$ is self-similar, as $x$ and $T$ enter only
through $\xi$. The integral in Eq.~(\ref{atT1}) can be expressed via the generalized hypergeometric
function \cite{Wolfram}, and we obtain $h(x,T\ll 1)=M \Phi(\xi)$, %where
\begin{eqnarray}
  \Phi(\xi)&=& _1F_3\left(-\frac{1}{4};\frac{1}{4},\frac{1}{2},\frac{3}{4};\frac{\xi^4}{256}\right)  \nonumber\\
  &+& \frac{\xi ^2\Gamma
   \left(\frac{1}{4}\right) \,_1\!F_3\left(\frac{1}{4};\frac{3}{4},\frac{5}{4},\frac{3}{2};\frac{\xi ^4}{256}\right)}{8
   \Gamma\left(\frac{3}{4}\right)}-\frac{\pi  |\xi| }{2 \Gamma\left(\frac{3}{4}\right)}. \label{Phiresult}
\end{eqnarray}
The graph of $\Phi(\xi)$ is shown in Fig. \ref{Phi}. The ``optimal" interface shape is oscillatory; it is very close to the one observed at $t=T$ on the bottom panel of Fig.~\ref{history}.

The long-time optimal height profile~(\ref{hTlong}) can also be obtained from a minimization of the equilibrium free energy (\ref{freen}), constrained by the conservation law $\int_{-L/2}^{L/2} h_T(x) \,dx = 0$.
%\begin{equation}\label{constraint}
%\int_{-L/2}^{L/2} h_T(x) \,dx = 0.
%\end{equation}
Indeed, introducing a Lagrange multiplier $\Lambda$, we arrive at a simple minimization problem for the Lagrangian
\begin{equation}\label{lagr}
L\left(h_T,h_T^{\prime} \right)= \left(h_T^{\prime}\right)^2-\Lambda h_T,
\end{equation}
where the primes denotes the $x$-derivative.
The Euler-Lagrange equation is $2 h_T^{\prime\prime}+\Lambda = 0$.
Its general solution is a parabola:
\begin{equation}\label{parabola}
h_T(x)=C_1+C_2 x-(1/4)\Lambda x^2.
\end{equation}
As $h_T(0)=M$, we obtain $C_1=M$.  Importantly, the solution cannot be smooth, because one cannot satisfy \emph{three} additional conditions -- the conservation law %(\ref{constraint})
and the
periodicity conditions $h_T(-L/2)=h_T(L/2)$ and $h_T^{\prime}(-L/2)=h_T^{\prime}(L/2)$ -- with only two arbitrary constants $C_2$ and $\Lambda$. The way out is to
allow a discontinuity of $dh/dx$ at $x=0$, so that the maximum at $x=0$ is non-analytic. In this case the coefficient $C_2$ does not vanish, and has the same magnitude but opposite signs at $x<0$ and $x>0$. This immediately leads to the optimal profile (\ref{hTlong}). This minimization problem has a simple mechanical analogy. Indeed, Eq.~(\ref{lagr}) describes the motion of a classical particle in a constant gravity field (directed upward), and the solution with the jump of the first derivative corresponds to a bounce of the particle from the ``floor" at $h=M$. Additional bounces (jumps in the first derivative) would cost more free energy,  so they are not allowed.

An extension of this theory to $d$ dimensions is straightforward. The short-time regime is a bit involved technically, but the equilibrium (long-time) regime is simple.  Here minimization of the free energy, constrained by the conservation law, yields a Gaussian distribution of $M$ with the variance
\begin{equation}\label{acnd}
\sigma_d(L) =\frac{L^{1-\frac{d}{2}}}{\sqrt{3}\cdot 2^{2-\frac{d}{2}}}.
\end{equation}
As one can see, $d=2$ is the critical dimension \cite{Siegert,Racz,Barabasi,Krug}.  The optimal surface at $t=T$ exhibits
a pyramid-like top.

In conclusion, we have employed the MFT to evaluate the probability that the stochastic Mullins-Herring interface with conserved noise first reaches a large given height at a specified time. We have also found the ``optimal" time history of the interface conditional on the first passage.  It would be interesting to apply the MFT to find the \emph{range} of the surface diffusion, that is to determine the joint distribution of the (large) maximum and minimum of the interface height at a given time.
The results of this work can be also applied to the stochastic Mullins-Herring equation with \emph{non-conserved} noise,  $\partial_{t}h=-\partial^4_x h+\xi (x,t)$, with the same $\xi$ as before. This equation describes,
in a moving frame, height fluctuations under particle deposition  \cite{Barabasi,Krug,Family}.  By differentiating this equation with respect to $x$, one again arrives at (the one-dimensional version of) Eq.~(\ref{mlnns}) but for the  local \emph{slope} of the interface  $s(x,t)=\partial_x h(x,t)$. This mapping yields the probability that the interface first develops a (large) given local slope, and the optimal interface slope history conditional on this event.

Finally, the exact solution of Eqs.~(\ref{eqh}) and (\ref{eqp}) was possible due to their linearity. Analogous MFT equations for nonlinear stochastic equations, such as the Kardar-Parisi-Zhang equation \cite{KPZ}, are harder to solve. The situation, however, is far from hopeless \cite{MKV}.

B.M. is grateful to Joachim Krug for useful discussions.  Financial support for this research was provided in part by
grant No.\ 2012145 from the United States-Israel Binational Science
Foundation (BSF).

\section*{Supplemental Material: Derivation of the MFT Equations}

The starting point of the derivation is the Langevin equation
\begin{equation}\label{mlnns10}
\partial_{t}h=-\partial^4_{x}h+\partial_{x}\xi(x,t),
\end{equation}
where $\xi$ is a delta-correlated Gaussian noise:
$$
\langle\xi(x_{1},t_{1})\xi(x_{2},t_{2})\rangle=\delta(x_{1}-x_{2})\delta(t_{1}-t_{2}).
$$
The problem is defined on the interval $|x|<L/2$ with periodic boundary conditions.  We are interested in the probability of transition from the initial state $h(x,t=0)=0$ to a given state $h(x,t=T)$ at a specified time $T$. Because of the conservation law, $h(x,T)$ must satisfy the condition $\int_{-L/2}^{L/2}h(x,T)dx=\int_{-L/2}^{L/2}h(x,0)dx=0$.
Let us introduce a potential $\psi(x,t)$ so that $\partial_{x}\psi=h(x,t)$. Now
Eq.~(\ref{mlnns}) becomes
\begin{equation}\label{mlnnspsi}
\partial_{t}\psi=-\partial^4_{x}\psi-\xi(x,t),
\end{equation}
and we can express
\begin{equation}\label{xi}
\xi(x,t)=-\partial_{t}\psi-\partial_{x}^4\psi.
\end{equation}
The Gaussian action is, therefore,
\begin{eqnarray}\label{actn}
S&=&\int_{0}^{T}dt\int_{-L/2}^{L/2}dx\frac{\xi^{2}(x,t)}{2} \nonumber \\
&=& \int_{0}^{T}dt\int_{-L/2}^{L/2}dx\frac{(\partial_{t}\psi+\partial_{x}^4\psi)^{2}}{2}.
\end{eqnarray}
Being interested in large deviations, we need to minimize this action with respect to the trajectory $\psi(x,t)$. The variation of the action is
\begin{equation}\label{variation}
\delta S=\int_{0}^{T}dt\int_{-L/2}^{L/2}dx(\partial_{t}\psi+\partial_{x}^4\psi)(\partial_{t}\delta\psi+\partial_{x}^4\delta\psi).
\end{equation}
Let us introduce the momentum density field $p(x,t)$, so that
\begin{equation}\label{px}
\partial_{x}p=-\partial_{t}\psi-\partial_{x}^4\psi.
\end{equation}
This follows
$$\partial_{tx}\psi=-\partial_{x}(\partial_{x}p+\partial_{x}^4\psi),$$
or
\begin{equation}\label{heqh}
\partial_{t}h=-\partial_{x}(\partial_{x}^3h+\partial_{x}p),
\end{equation}
one of the Hamilton equations. Now we can rewrite the variation (\ref{variation}) as follows:
$$
\delta s=-\int_{0}^{T}dt\int_{-L/2}^{L/2}dx\,\partial_{x}p(\partial_{t}\delta\psi+\partial_{x}^4\delta\psi).
$$
After several integrations by parts, we obtain the Euler-Lagrange equation
\begin{equation}\label{heqp}
\partial_{xt}p=\partial_{x}^5 p
\end{equation}
which yields the second Hamilton equation of the MFT formalism:
\begin{equation}\label{peqh}
\partial_{t}p=\partial_{x}^4p.
\end{equation}
The boundary terms in time and in space, resulting from the integrations by parts, all vanish because we specified fixed states at $t=0$ and $t=T$, and because of the periodic boundary conditions in $x$.

\end{document}